\documentclass[11pt]{article}
\usepackage{style}
\usepackage{natbib}
\usepackage{fullpage}
\usepackage{tikz}

\usepackage{suffix}
\WithSuffix\newcommand\citet*{\cite}
\Crefname{algocf}{Algorithm}{Algorithms}

\newcommand{\halt}{\mathsf{Halt}}
\newcommand{\yes}{\mathsf{Yes}}
\newcommand{\no}{\mathsf{No}}

\title{Differentially Private All-Pairs Min Cut Values}
\author{
Anders Aamand\thanks{Rice University. \texttt{aa561@rice.edu}.
} \and 
Justin Y. Chen\thanks{Massachusetts Institute of Technology. \texttt{justc@mit.edu}. Supported by an NSF
Graduate Research Fellowship under Grant No.\ 17453.} \and 
Mina Dalirrooyfard\thanks{Morgan Stanley. \texttt{minad@mit.edu}.} \and 
Slobodan Mitrovi{\'c}\thanks{UC Davis. \texttt{smitrovic@ucdavis.edu}. Supported by the NSF Faculty Early Career Development Program No.~2340048.} \and 
Sandeep Silwal\thanks{UW-Madison. \texttt{silwal@cs.wisc.edu}.} \and 
Yinzhan Xu\thanks{UC San Diego. \texttt{xyzhan@ucsd.edu}. Supported by NSF HDR TRIPODS Phase II grant 2217058 (EnCORE Institute). }
}
\date{}

\begin{document}

\maketitle

\begin{abstract}
We study the problem of privately releasing the min cut values of all pairs of vertices in a given $n$-node weighted graph, under the edge-neighboring model. Straightforward application of advanced composition yields an approximate DP algorithm that outputs all min cut values up to $\tilde{O}(n)$ additive error. Prior works [Dalirrooyfard, Mitrovi{\'c} and Nevmyvaka, 2023; Aamand et al., 2026] have established that $\tilde{\Theta}(n)$ is the correct bound if the algorithm also needs to output cuts that realize the approximate min cut values. On the other hand, simply outputting the values could be an easier problem, but prior algorithms have been stuck on the $\tilde{O}(n)$ bound.

In this work, we give an approximate DP algorithm outputting all-pair min cut values with $\tilde{O}(\sqrt{n})$ additive error, beating the linear error barrier. Our algorithm applies the sparse vector technique [Dwork and Roth, 2014] to maintain a hierarchical decomposition of the vertices based on the min cut structure. We also prove an $\Omega(\log^{3/2} n)$ lower bound for this problem. 
\end{abstract}

\newpage
\section{Introduction}
% shortest path values only
% previous lower bound
% more value-only results?
% On DP for adaptively solving search problems via sketching?
% Hierarhical clustering if you have values only?
% k-edge connectivity
% vertex importance
Given a weighted graph\footnote{We let $|V| = n$ and we require the weights to be non-negative.} $G = (V, E, w)$, a vertex partition $(S, V \setminus S)$ is called a cut, and the sum of weights of the edges with one endpoint in $S$ and another endpoint in $V \setminus S$ is called the \emph{value} of that cut.
Recently, there has been increasing interest in understanding cut problems in the context of differential privacy, including global min cut~\cite{gupta2010differentially}, min $s$-$t$ cut~\cite{dalirrooyfard2023nearly,chandra2026differentially}, all-pairs min $s$-$t$ cuts~\cite{aamand2025differentially}, cut sparsifier~\cite{gupta2012iterative,eliavs2020differentially,liu2024optimal,aamand2025breaking}, and more~\cite{hsu2014privately,dalirrooyfard2023nearly,chandra2026differentially}. In these works and also in our paper, the model of neighboring datasets is edge-neighboring: two neighboring graphs are those whose total edge weights differ by at most one, and in a single edge.

We consider one of the simplest problems: min $s$-$t$ cut. This problem admits an interesting behavior depending on what exactly we wish to output. For a fixed pair $(s,t)$, it is known how to output an entire $s$-$t$ cut\footnote{By this we mean a vertex partition with $s$ and $t$ on opposite parts.} with an $\eps$-DP algorithm, whose cut value in the underlying graph differs by $\tilde{O}(n/\eps)$ from the actual min $s$-$t$ cut \cite{dalirrooyfard2023nearly}. The $\tilde{O}(n)$ additive error bound is also known to be tight \cite{dalirrooyfard2023nearly}. Interestingly, the same $\tilde{O}(n/\eps)$ bound can be achieved even if one needs to privately output an approximate min $s$-$t$ cut for \emph{every pair} $s, t \in V$ \cite{aamand2025differentially}. 
% For some of these problems, e.g., for releasing a min $s$-$t$ cut or even all min $s$-$t$ cut pairs, nearly matching upper and lower bound on approximation are known~\cite{dalirrooyfard2023nearly,aamand2026differentially}.

On the other hand, if the goal is instead to output the (approximate) cut values, one can potentially do much better. For example, to output a single min $s$-$t$ cut \emph{value} privately, we only need to naively add $\Lap(1/\eps)$ noise to the true min $s$-$t$ cut value, resulting in $O(1/\eps)$ additive error (in expectation), which is much smaller than the $\tilde{O}(n/\eps)$ bound. 
However, it seems more challenging to output the values of all min $s$-$t$ cut with lower additive error than privately releasing the cuts themselves. 
Namely, since there are $n^2$ values to output, we can apply advanced composition to achieve additive error $\tilde{O}(n)$ (with approximate DP), which matches the error bound of \cite{aamand2025differentially}, but is significantly weaker since the algorithm in \cite{aamand2025differentially} is able to output the cuts themselves. This has prompted \cite{aamand2025differentially} to raise the natural open question of
whether there is an algorithm that releases all-pairs min-cut values with error polynomially
smaller than $\tilde{O}(n)$; the distinction between releasing cuts and releasing only their
values has also been noted in \cite{dalirrooyfard2023nearly, chandra2026differentially}.

% However, when it comes to reporting \emph{values} of the cuts only, the situation is much less understood. 
% On the one hand, $s$-$t$ cut can be output in a DP way with $O(n/\eps)$ additive approximation~\cite{dalirrooyfard2023nearly}, while the min $s$-$t$ cut value can be output with essentially exponentially smaller additive approximation; by adding $\Lap(1/\eps)$ to the underlying min $s$-$t$ cut value.
% On the other hand, outputting $\binom{n}{2}$ min $s$-$t$ cut values can be done with $\tO(n / \eps)$ additive approximation, which directly follows from the result on outputting all such cuts themselves~\cite{aamand2026differentially}. 
% It is unknown whether a lower additive approximation can be achieved. 
In this work, we answer this open question in the affirmative, and show how to output all min $s$-$t$ cut values with quadratically lower additive error than outputting the cuts themselves.
% Perhaps bring up advanced decomposition in a footnote.
\begin{theorem}\label{thm:main}
    Let $\eps,\delta,\beta>0$. There exists an $(\eps,\delta)$-DP algorithm which for any non-negatively weighted $n$ vertex graph $G=(V,E,w)$ outputs values $\tilde \lambda(s,t)$ for each $s,t\in V$. If the value of the $s$-$t$ min-cut in $G$ is $\lambda(s,t)$, then with probability at least $1-\beta$ it holds simultaneously over all $s,t\in V$ that  
\[
|\tilde \lambda(s,t)-\lambda(s,t)|=O\left(\frac{\sqrt{n\log (1/\delta)}}{\eps}(\log n+\log(1/\beta))\right).
\]
\end{theorem}

We also give a (much weaker) lower bound, showing that releasing all min $s$-$t$ cut values is quantitatively harder than releasing a single min-cut value: \cref{cor:apmc-value-lower-bound} shows that, for a natural range of $\eps$ and $\delta$, every $(\eps, \delta)$-DP mechanism which outputs all cut values from a \emph{fixed} source node $s$ must incur additive error $\Omega\left(  \frac{\log^{3/2}n}{\max\{\eps, \delta\}}\right)$ on some cut value. In contrast, prior works of \cite{dalirrooyfard2023nearly, aamand2025differentially} show that both tasks are equally efficient (both have $\tilde{\Theta}_{\eps, \delta}(n)$ additive error), if one is also required to release cuts certifying the min-cut values. Notably, our lower bound holds for an easier problem since we fix one end point. 

\paragraph{Open Problems.}
Given the exponential gap between our upper and lower bounds, a natural open problem is to improve the bounds in either direction towards discovering the asymptotically optimal bounds.
It is also interesting to ask for improved bounds for the problem of releasing all min $s$-$t$ cut values for a fixed public node $s$, the ``single-source'' version of the problem studied in this work. For this simpler problem, naively applying advanced composition already gives $\tilde{O}_{\eps, \delta}(\sqrt{n})$ dependence, and we ask if this bound can be improved.

\section{Preliminaries}

Let $G=(V, E, w)$ be a weighted graph, where we throughout this paper assume that $E=\binom{V}{2}$; we interpret edges of weight zero as non-edges.
For $u, v \in V$, we use $\lambda_G(u, v)$ to denote the value of a min cut between $u$ and $v$ in the graph $G$. We recall some basic DP definitions.

\begin{definition}[Edge-Neighboring Graphs]
Graphs $G=(V,E,w)$ and $G'=(V,E,w')$ are called \emph{edge-neighboring} if there is $e\in E$ such that $|w(e)-w'(e)|\le 1$ and for all $e'\in E\setminus \{e\}$,  $w(e')=w'(e')$. 
\end{definition}

\begin{definition}[Differential Privacy \cite{dwork2006differential}]\label{def:dp} 
A randomized algorithm $\mathcal{A}$ is $(\eps,\delta)$-DP if for any neighboring graphs $G$ and $G'$ and any set of possible outcomes $O$ of $\mathcal{A}$ it holds $\prob{\mathcal{A}(G)\in O} \le e^{\eps}\prob{\mathcal{A}(G')\in O}+\delta$. When $\delta=0$, algorithm $\mathcal{A}$ is called \emph{pure DP}, or $\eps$-DP.
\end{definition}

Now we describe the version of the sparse vector technique (SVT) that we require for our main algorithm for releasing all pairwise min-cut values under differential privacy. 
Let $\mathcal X$ be an input space equipped with a neighboring
relation $\sim$. For a scalar query $q:\mathcal X\to\mathbb R$, define its
global $\ell_1$-sensitivity by
\[
  \Delta(q)
  := \sup_{x,x'\in \mathcal{X}, x\sim x'} |q(x)-q(x')|.
\]
\paragraph{Adaptive threshold queries.}
We employ an adaptive version of SVT that allows us to approximately answer a sequence of questions of the form ``\textit{Is $q_i(x) \ge T_i$?}'' for a sequence of scalar queries $q_1,\dots, q_K$ each of $\ell_1$ sensitivity at most $\Delta$ and a sequence of thresholds $T_1,\dots,T_K$. Adaptivity means that the algorithm can choose $q_i$ and $T_i$ based on answers to previous questions.
At round $i$, after observing the preceding transcript $h_{i-1}\in\{\mathsf{Yes},\mathsf{No}\}^{i-1}$, the algorithm chooses a new query $q_i\colon\mathcal X\to\mathbb R$ and a threshold
$T_i\in\mathbb R$. 
SVT roughly states that we are able to get accurate answers up to additive error scaling with $\approx \sqrt{c}$, where $c$ is the total number of $\mathsf{Yes}$ answers we will get.

% \textcolor{red}{add intuition for the theorem below}
% \textcolor{blue}{Slobo: I will take this intuition part.}
% Before we formally state the SVT theorem, we provide a brief intuition behind it.
% SVT allows designing a private mechanism for answering a sequence of thresholding queries while ``paying'' in privacy only for the ``Yes'' answers; the cost for the ``No'' answers is almost negligible, only logarithmic in their number.
% Precisely, if we aim to answer at most $c$ ``Yes'' queries, then SVT enables to do so in a DP manner by incurring an additive error with only $\sqrt{c}$ dependence. This dependence in the theorem below is given by $\alpha$.
% 

\begin{restatable}[SVT]{thm}{svt}
\label{thm:svt}
Consider any $\eps \in (0,1)$, $\delta, \Delta, \beta > 0$, and $c \in \mathbb{N}$.
Consider a private dataset $D$ and a sequence $q_1, \ldots, q_k$ of (adaptively chosen) sensitivity $\Delta$ queries along with a sequence of thresholds $T_1, \ldots, T_k$.
Let
\[
\alpha = \frac{16\Delta \sqrt{2c\ln(1/\delta)} (\log k + \log(2c/\beta)) }{\eps}
\]
There exists an $(\eps, \delta)$-DP algorithm which outputs a sequence $a_1, \ldots a_k$ with $a_i \in \{\yes,\no, \halt\}$ such that the following holds with probability $1-\beta$ across all timesteps $i \in \{1, \ldots, k\}$
\begin{itemize}
    \item If $a_i = \yes$, then $q_i(D) \geq T_i - \alpha$.
    \item If $a_i = \no$, then $q_i(D) \leq T_i + \alpha$.
    \item If $a_i = \halt$, then $|\{a_j : j < i \text{ and } a_j = \yes\}| = c$ (the algorithm has already answered $\yes$ $c$ times).
\end{itemize}
\end{restatable}

This version of the SVT theorem which we use slightly differs from a standard presentation in \cite{DworkRoth2014} though follows immediately from the same techniques, we defer its proof to \cref{sec:svt}.

% \begin{theorem}[{\cite[Theorems~3.25--3.26]{DworkRoth2014}}]
% Let $K,c\ge 1$, $0<\varepsilon<1$, $0<\delta\le e^{-1}$, and $0<\beta<1$. Furthermore, let
% \[
%   \alpha
%   =\frac{1}{\varepsilon}  \sqrt{512c\ln\frac1\delta} \left( \ln K+\ln\frac{2c}{\beta}\right).
% \]
% Let $(q_1,T_1),\dots,(q_K,T_K)$ be an adaptively chosen sequence of sensitivity one scalar queries to a data set $x\in \mathcal{X}$ such that $|\{i\in [K]\mid q_i(x)\geq T_i-\alpha\}|<c$. There is an $(\varepsilon,\delta)$-DP interactive mechanism that with probability at least $1-\beta$ produces answers $(a_i)_{i=1}^K\in\{\mathsf{Yes},\mathsf{No}\}^K$ such that simultaneously for every $i\in[K]$,
% \[
%   \begin{aligned}
%   a_i=\mathsf{Yes}
%     &\quad\Longrightarrow\quad
%       q_i(x)\ge T_i-\alpha,\\
%   a_i=\mathsf{No}
%     &\quad\Longrightarrow\quad
%       q_i(x)\le T_i+\alpha.
%   \end{aligned}
% \]
% \end{theorem}
% \paragraph{Remark:} We note that~\cite{DworkRoth2014} considers the case where all $T_i$ are the same. However, the case where the $T_i$ can be different follows as a direct corollary after applying the fixed threshold version to the modified scalar queries $\tilde q_i(x)=q_i(x)-T_i$ with common threshold 0.

\section{Algorithm for DP All-Pairs Min-Cut Values}
In this section we prove \cref{thm:main}. 
For simplicity, we first assume that for the graph $G=(V,E,w)$, $w$ is bounded by a polynomial in $n$. 
At the end of the section, we discuss how to handle arbitrary non-negative edge weight. 
We start with the following observation.

\begin{observation}\label{lem:cut-transitivity}
    Let $s,t,v$ be three arbitrary vertices in the graph, and $T \ge 0$ be a real number. If $\lambda_G(s,t) \ge T$ and $\lambda_G(s,v) \ge T$, then $\lambda_G(t,v) \ge T$ as well.
\end{observation}
\begin{proof}
    Suppose $\lambda_G(t,v)$ is $T'<T$. Consider the two components of a Min-$t$-$v$-Cut: $A_t$ and $A_v$. If $s\in A_t$, then $(A_t, A_v)$ is an $s$-$v$-Cut as well, and so $\lambda_G(s,v)$ is at most $T'<T$, a contradiction. Similarly, if $s\in A_v$, then $(A_t, A_v)$ is a $s$-$t$-Cut as well and so $\lambda_G(s,t)$ is at most $T'<T$, a contradiction. So $\lambda_G(t,v)$ must be at least $T$.
\end{proof}
As in~\cref{thm:svt}, we define
\[
\alpha = \frac{16 \sqrt{2n\ln(1/\delta)} (\log k + \log(2n/\beta)) }{\eps}.
\]
Here, $k$ is an upper bound on the number of SVT scalar queries our algorithm will make. 
We later upper-bound $k$ by $n^{O(1)}$. In particular, \cref{thm:svt} guarantees that under $(\eps,\delta)$-DP, with probability at least $1-\beta$, we can answer up to $k$ scalar queries of sensitivity at most one within error $\alpha$ as long as the number of $\yes$ answers is at most $n$. 
Our scalar functions are of the form $q_{C_1,C_2}$ for disjoint subsets $C_1,C_2\subset V$ where
\[
    q_{C_1,C_2}(G)=\max_{u\in C_1,v\in C_2} \lambda_G(u,v). 
\]
Clearly $q_{C_1,C_2}$ has sensitivity at most one under the notion of edge-neighboring graphs. 

%for distinct nodes $s,t$, where $q_{s,t}(G)$ is the size of an \minstcut of $G$. Clearly $q_{s,t}$ has sensitivity at most one under the notion of edge-neighboring graphs. 
Our algorithm consists of $\poly(n)$ many steps each with a different cut threshold. At each step $i$ with threshold $T_i$, the vertices will be decomposed into components, such that if $s$ and $t$ are in the same component, $\lambda_G(s,t)$ is at least $T_i-\alpha$, and if they are in different components, $\lambda_G(s,t)$ is smaller than $T_i+\alpha$. At step $i+1$, we lower the threshold and coarsen this decomposition.

At step $0$ we start with a high threshold $T_0$ such that the $\lambda_G(s,t)<T_0-\alpha$ for all $s,t\in V$ and having all components be singletons. By the assumption that the edge weights are polynomially bounded, it suffices to set $T_0=\poly(n)$. Going from step $i$ to step $i+1$, we let $T_{i+1} = T_i-\alpha$. 

Suppose at the beginning of step $i$, we have $t_i$ components $C_1,\ldots,C_{t_i}$. We take two arbitrary components $C_j,C_{j'}$ and query if $q_{C_j,C_{j'}}(G)$ is at least $T_{i}$. If the answer is $\yes$, we merge $C_j$ and $C_{j'}$. We continue until we coarsen the decomposition to $t_{i+1}\le t_i$ many components where for each pair of current components $C,C'$ we have performed the above threshold query $q_{C,C'}(G)$ with answer $\no$.

\begin{lemma}
    At step $i$, we perform $O(t_i^2)$ queries. If the number of $\yes$ queries during step $i$ is $y_i$, then $t_{i+1} = t_i-y_i$, i.e. the number of components at step $i+1$ is the number of components at step $i$ minus the number of YES queries.
\end{lemma}
\begin{proof}
During step $i$, two components can merge at most $t_i-1$ times, so there are at most $2t_i-1$ distinct components in total that can ever be part of a query, including the $t_i$ initial components and the components created by merging. Since each pair of components is queried at most once during step $i$, the total number of queries during step $i$ is at most $\binom{2t_i-1}{2}=O(t_i^2)$.
Each time we get a YES, the number of components goes down by one which proves the second statement.
\end{proof}
\begin{corollary}\label{corollary:few-yes-queries}
    The total number of $\yes$ queries the algorithm asks is at most $n-1$ and the total number of queries is $\poly(n)$.
\end{corollary}
\begin{proof}
The first statement follows directly from the previous lemma, and the second statement follows since we assumed the weights of $G$ and hence $T_0$ is polynomially bounded. 
\end{proof}
The next lemma shows that we can approximately recover the values of the min-cuts from the steps $i$ at which components merged.
\begin{lemma}\label{lem:when-components-merge}
With probability at least $1-\beta$, the following holds simultaneously for every step $i\geq 1$: if $C$ and $C'$ are two components that merge at step $i$, then, for every $s\in C$ and $t\in C'$, $\lambda_G(s,t)$ is in $[T_i-\alpha, T_{i-1}+\alpha]$.
\end{lemma}
\begin{proof}
Let $\mathcal{E}$ be the event that all threshold queries are answered within error $\alpha$. Since the total number of $\yes$ queries is at most $n-1$ by~\cref{corollary:few-yes-queries}, it follows from~\cref{thm:svt} and the way we set $\alpha$ that $\Pr[\mathcal{E}]\geq 1-\beta$. It thus suffices to show that if $\mathcal{E}$ occurs, $\lambda_G(s,t)$ is in $[T_i-\alpha, T_{i-1}+\alpha]$ deterministically. 
We use induction on $i$. For $i=0$, there is nothing to prove. However, note that since all min-cuts in the graph are smaller than $T_0-\alpha$, under $\mathcal{E}$, no components are merged during step $i=0$. Assume that $i>0$ and that the result holds for smaller values of $i$.

We first prove the upper bound. Let $C_0$ and $C_0'$ be the components containing $s$ and $t$, respectively, at the end of step $i-1$. Since $C_0$ and $C_0'$ did not merge during step $i-1$, the query $q_{C_0,C_0'}$ with threshold $T_{i-1}$ was answered with $\no$. Under $\mathcal{E}$, this implies that for every pair of nodes $u\in C_0$ and $v\in C_0'$, $\lambda_G(u,v)\leq T_{i-1}+\alpha$, and in particular $\lambda_G(s,t)\leq T_{i-1}+\alpha$.

It remains to prove that $\lambda_G(s, t) \geq T_i-\alpha$. We prove that, throughout step $i$, any two vertices in the same component have min-cut value at least $T_i-\alpha$. This holds at the beginning of step $i$ by the induction hypothesis and the fact that the thresholds are decreasing. Now suppose that $C$ and $C'$ were merged during step $i$. Under $\mathcal{E}$, this implies that $q_{C,C'}(G)\geq T_i-\alpha$ since the query was answered with a $\yes$. Consequently, there exist $u\in C$ and $v\in C'$ such that $\lambda(u,v)\geq T_i-\alpha$. Inductively, all pairs of vertices within $C$, and all pairs within $C'$, have min-cut values at least $T_i-\alpha$.
Since having min-cut at least $T_i-\alpha$ is an equivalence relation by~\cref{lem:cut-transitivity}, it follows that $\lambda_G(s, t) \geq T_{i}-\alpha$, as desired. 

%Suppose for contradiction that $\lambda(s, t) \geq T_{i-1}+\alpha$. Let $C_0$ and $C_0'$ be the components that contained $s$ and $t$ at the end of step $i-1$. By the induction hypothesis, for distinct $x,y\in C_0$, it holds that $\lambda(s,t) \geq T_{i-1}-\alpha$ and the same statement holds for $x',y'\in C_0'$. By~\cref{lem:cut-transitivity}, having min-cut at least $T_{i-1}$ is an equivalence relation on nodes, so it follows that for all $u\in C_0$ and $v\in C_0'$, $\lambda(u, v) \geq T_{i-1}$. On the other hand, since $C_0$ and $C_0'$ did not merge during step $i-1$, there does not exist nodes $u\in C_0$ and $v\in C_0'$ such that $\lambda(u, v) \ge T_{i-1}$, a contradiction. 
\end{proof}
We can now prove our main result

\begin{proof}[Proof of~\cref{thm:main}]
We run the algorithm described above. As weights are nonnegative, when $i$ is large enough that $T_{i-1}<-\alpha$, $t_i=1$ with probability at least $1-\beta$, so there is just one component on all $n$ nodes left. For any pair of nodes $s,t$, we note the step $i$ in which the components of $s$ and $t$ are merged and our estimate of $\lambda_G(s,t)$ is simply $T_i$. By~\cref{lem:when-components-merge}, with probability at least $1-\beta$, the error is
\[
|T_i-\lambda_G(s,t)|\leq 2\alpha.
\]
Using~\cref{corollary:few-yes-queries}, the number of queries $k$ is polynomial in $n$, so it follows that the error $2\alpha$ is bounded by 
\[
O\left(\frac{\sqrt{n\log (1/\delta)}}{\eps}(\log n+\log(1/\beta))\right).
\]
The algorithm is $(\eps,\delta)$-DP as the merge history and hence the estimates are post-processing of the $(\eps,\delta)$-DP SVT answers.
\end{proof}

\paragraph{Handling Unbounded Weights.}
While we proved our result for polynomially bounded weight functions, there is a simple trick to obtain an algorithm with no such assumption. The idea is to use half the privacy budget to first obtain coarse estimates of all pairwise min-cut values. This can be done by adding independent Laplace noise $\Lap(2/\eps)$ to each edge weight of $G$ and computing the min-cut values on the resulting graph. With probability at least $1-\beta/2$, no edge changes weight by more than $O(\log(n/\beta)/\eps)$, so all cut values are preserved within $O(n^2\log(n/\beta)/\eps)$. 

We use the second half of the privacy budget to simulate the algorithm behind~\cref{thm:main}. However, since we have coarse bounds on the min-cut values, instead of running the threshold queries for every step $i$, we only need to consider steps $i$ such that $T_i$ is within $O(n^2\log(n/\beta)/\eps+\alpha)$ of \emph{some} min-cut value of the graph. Since there are at most $n-1$ different min-cut values in any non-negatively weighted graph \cite{gomory1961multi}, the number of steps $i$ we need to consider is therefore
\[
O\left((n-1) \cdot \frac{n^2\log(n/\beta)/\eps+\alpha}{\alpha}\right) 
= O\left(\frac{n^3\log(n/\beta)}{\alpha\eps}+n\right)
=O\left(\frac{n^{5/2}}{\sqrt{\log (1/\delta)}}+n\right)=O(n^{5/2}),
\]
and the number of queries $k$ is at most a polynomial factor in $n$ larger. We thus obtain that $\log k=O(\log n)$. Setting the failure probability of the second algorithm to $\beta/2$, a union bound gives that the final algorithm fails with probability at most $\beta$, and the final error remains unchanged within constant factors.

\section{A Lower Bound for Releasing All-Pairs Min-Cut Values}
In this section, we prove an $\Omega_{\eps,\delta}(\log^{3/2} n)$ error lower bound for releasing all-pairs min-cut values. The lower bound holds for an easier problem of releasing all the min-cut values from a \emph{single fixed} source.  The main idea is a reduction from the well studied problem of releasing prefix sums, which is classical in differential privacy \cite{dwork2010differential, chan2011private}. We recall its definition below.
\begin{definition}[DP Prefix-Sum Problem]\label{def:prefix-sums}
For a string $x=(x_1,\ldots,x_d)\in\{0,1\}^d$ and $k \in [d]$, define its prefix sums as $p_k(x)=\sum_{j=1}^k x_j$. Two strings $x$ and $x'$ are called neighboring if they differ in
exactly one coordinate. The DP prefix-sum problem asks to output all
the values $p_1(x),\ldots,p_d(x)$ under this neighboring definition.
\end{definition}

A recent work proves the following
lower bound \cite{bairaktari2026binary}. Importantly for our setting, their lower bound holds even in the offline model, in which the mechanism receives the entire
string before producing its output.

\begin{theorem}[\cite{bairaktari2026binary}]
\label{thm:prior_cc_lb}
There is a sufficiently small constant $C>0$ such that, for any
\[
    d^{-1+\Omega(1)}<\eps<1
    \qquad\text{and}\qquad
    0<\delta<C,
\]
every $(\eps,\delta)$-DP mechanism
$\mathcal{A}\colon\{0,1\}^d\to\mathbb{R}^d$ for the offline prefix-sum
problem satisfies
\[
    \max_{x\in\{0,1\}^d}
    \mathbb{E}\left[
        \left\|
            \mathcal{A}(x)
            -
            \bigl(p_1(x),\ldots,p_d(x)\bigr)
        \right\|_\infty
    \right]
    =
    \Omega\left(
        \frac{\log^{3/2} d}{\max\{\eps,\delta\}}
    \right).
\]

\end{theorem}
\paragraph{Remark:} Here, the condition $d^{-1+\Omega(1)}<\eps$ means that
$\eps\ge d^{-1+\gamma}$ for some fixed constant $\gamma>0$. The
implicit constant in the lower bound may depend on $\gamma$.

We now give a reduction from the problem of
\cref{def:prefix-sums} to releasing all min-cut values from a single
fixed source. The main idea in the lemma is to encode prefix sums as min cuts from a single source node, using a modified path graph; we refer to \cref{fig:lowerbound_construction}.

\begin{lemma}\label{lem:prefix_reduction}
For every $x \in \{0, 1\}^d$, there is a simple undirected weighted graph $G_x$ on $d+2$ vertices, with
a source $s$ and vertices $v_0,\ldots,v_d$, such that
\begin{equation}\label{eq:prefix_relation}
     \lambda_{G_x}(s,v_{k-1})=d+1+p_k(x)
    \qquad\text{for every }k\in[d].
\end{equation}
If $x$ and $x'$ differ in one
coordinate, then $G_x$ and $G_{x'}$ differ in the weight of one edge by
$1$. 
\end{lemma}

\begin{proof}  Given a string $x \in \{0, 1\}^d$, we construct the corresponding graph $G_x$ as follows. This is also shown in \cref{fig:lowerbound_construction}. Let the vertices be $ v_0,v_1,\ldots,v_{d+1}$ and set the source as $s=v_{d+1}$. First connect the vertices $v_i$ in a path with all edge weights $W:= d+1$. That is, 
for each $i\in[d+1]$, add the path edge $e_i=\{v_{i-1},v_i\}$ with weight $W$. This is independent of the string $x$. Now to incorporate the string, for each $j\in[d]$, add the edge  $\{v_{j-1},s\}$ with weight $x_j$. For example, when $d=4$, the construction is shown in \cref{fig:lowerbound_construction}. 

\begin{figure}[!htbp]
\begin{center}
\begin{tikzpicture}[
    x=1cm,
    y=1cm,
    vertex/.style={circle,fill=black,inner sep=2pt},
    every node/.style={font=\small}
]
    \node[vertex] (v0) at (0,0) {};
    \node[vertex] (v1) at (1.6,0) {};
    \node[vertex] (v2) at (3.2,0) {};
    \node[vertex] (v3) at (4.8,0) {};
    \node[vertex] (v4) at (6.4,0) {};
    \node[vertex] (v5) at (8.0,0) {};

    \node at (0,-0.45) {$v_0$};
    \node at (1.6,-0.45) {$v_1$};
    \node at (3.2,-0.45) {$v_2$};
    \node at (4.8,-0.45) {$v_3$};
    \node at (6.4,-0.45) {$v_4$};
    \node at (8.0,-0.45) {$s=v_5$};

    \draw[thick] (v0)--node[below=2pt] {$W$} (v1);
    \draw[thick] (v1)--node[below=2pt] {$W$} (v2);
    \draw[thick] (v2)--node[below=2pt] {$W$} (v3);
    \draw[thick] (v3)--node[below=2pt] {$W$} (v4);
    \draw[thick] (v4)--node[below=2pt] {$W$} (v5);

    \draw[dashed]
        (v0) to[bend left=24]
        node[pos=.1,above] {$x_1$} (v5);
    \draw[dashed]
        (v1) to[bend left=28]
        node[pos=.1,above] {$x_2$} (v5);
    \draw[dashed]
        (v2) to[bend left=34]
        node[pos=.1,above] {$x_3$} (v5);
    \draw[dashed]
        (v3) to[bend left=44]
        node[pos=.1,above] {$x_4$} (v5);

\end{tikzpicture}
\end{center}
\caption{Example of the lower bound gadget.}
    \label{fig:lowerbound_construction}
\end{figure}
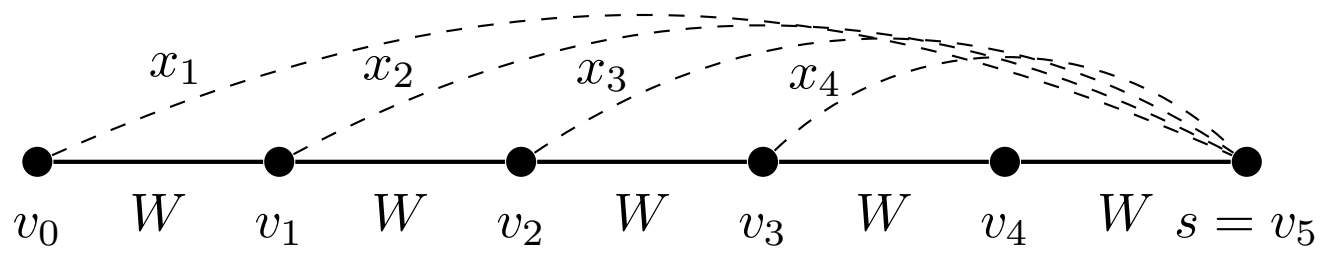
It is clear that for two neighboring strings $x, x'$ in the sense of \cref{def:prefix-sums}, their corresponding graphs differ in exactly one edge by weight one. 

Now the important claim is that the min cuts from the source $s$ are exactly the prefix sums of the string $x$, up to an additive shift $W$ which is known. Indeed, consider the min-cut between $s$ and node $v_i$ for some $i \le d-1$. Let's imagine the vertices as linearly ordered as shown in \cref{fig:lowerbound_construction}. Due to our choice of $W$, we never want to cut more than one weight-$W$ edge and it is clear that at least one weight-$W$ edge must be cut. It is also clear that a weight-$W$ edge \emph{after} $v_i$ must be cut, otherwise there is a path of weight-$W$ edges from $v_i$ to $s$. If we cut some  weight-$W$ edge adjacent to $v_i'$ for $i' \ge i$, then we must also cut all the edges corresponding to $x_1, \cdots, x_{i'+1}$ (otherwise there is a path from $s$ to $v_i$), and these additional edges we cut incur cost which is equal to the prefix sum $p_{i'+1}(x)$. But since prefix sums are non-decreasing, it never hurts to take $i' = i$, in which case we get the prefix sum $p_{i+1}(x)$, and shifting the index by $1$ proves \eqref{eq:prefix_relation}.
\end{proof}
Consequently, any $(\eps,\delta)$-DP mechanism for releasing all
min-cut values of $G_x$ from the source $s$ gives an
$(\eps,\delta)$-DP mechanism for releasing all prefix sums of $x$.
Indeed, if we have min cut estimates $\tilde{\lambda}(s,v_{k-1})$, we output the prefix sum estimate $
    \tilde{\lambda}(s,v_{k-1})-(d+1).$ This reduction preserves both differential privacy and additive
$\ell_\infty$ error. This observation along with the reduction of \cref{lem:prefix_reduction} and the prior result of \cref{thm:prior_cc_lb} immediately proves the following corollary.

\begin{corollary}\label{cor:apmc-value-lower-bound}
There is a sufficiently small constant $C>0$ such that, for any
$n^{-1+\Omega(1)}<\eps<1$ and $0<\delta<C,$ every $(\eps,\delta)$-DP mechanism that, on an $n$-vertex simple
undirected weighted graph with a fixed public source $s$, releases
estimates $\tilde{\lambda}_G(s,v)$ for every $v\neq s$ must satisfy
\[
    \max_G
    \mathbb{E}\left[
        \max_{v\in V(G)\setminus\{s\}}
        \left|
            \tilde{\lambda}_G(s,v)-\lambda_G(s,v)
        \right|
    \right]
    =
    \Omega\left(
        \frac{\log^{3/2} n}{\max\{\eps,\delta\}}
    \right).
\]
\end{corollary}
There are several interesting remarks to make. The first is that the lower bound holds for the easier problem of releasing all min cut values from a single fixed source, rather than APMC values. It is an interesting question to determine if this is the right answer for the fixed source problem. The second is that the result shows a separation between releasing one fixed min cut value and releasing APMC values. For a single min cut, one can obtain expected error $O(1/\eps)$ using the Laplace mechanism, whereas APMC values requires error that must grow with the number of vertices. Lastly, the lower bound shows that APMC is a ``harder'' problem than privately releasing all vertex degrees of a graph (which are perhaps the most natural overestimates of the true min cut values), since degrees can be released with $O_{\eps}(\log n)$ error.

\section*{Declaration of AI Use}
The lower bound of \cref{cor:apmc-value-lower-bound} was derived by ChatGPT 5.6 Pro when the authors prompted it with a prior weaker lower bound based on the packing argument. The writing was heavily edited by the authors and the authors assume responsibility for all content.

% \section*{Acknowledgements}
% Justin Y. Chen was supported by an NSF Graduate Research Fellowship under Grant No. 17453.
% Slobodan Mitrovi\'c was supported by the NSF Faculty Early Career Development Program No. 2340048. Yinzhan Xu was supported by NSF HDR TRIPODS Phase II grant 2217058 (EnCORE Institute).

\bibliographystyle{alpha}
\bibliography{bib}

\appendix

\section{Proof of SVT Theorem}
\label{sec:svt}

We give a proof of a slightly non-standard utility theorem for the Sparse Vector Technique (SVT).
This follows immediately from standard analysis found in \cite{DworkRoth2014} but is not precisely the theorem stated there.
For completeness, we include the proof here.

\paragraph{Background from \cite{DworkRoth2014}.}
Consider the AboveThreshold algorithm (Algorithm 1 in \cite{DworkRoth2014}), which takes in a private dataset $D$, a privacy parameter $\eps$, a sequence $f_1, \ldots, f_k$ of (adaptively chosen) sensitivity $1$ queries, and a threshold $T$.
After receiving each query, the algorithm either outputs $\no$ and continues to the next query or outputs $\yes$ and halts, it answers no more queries after this point.
Let $a_i$ be the output after receiving query $f_i$.
% \textcolor{blue}{Slobo: What exactly is meant by the previous sentence, the part ``if the algorithm has already halted''? Are you now switching to multiple halts allowed?}
In Theorem 3.23 of their work, the authors prove that this algorithm $(\eps, 0)$-DP.

\begin{lemma}
\label{lem:above-dp}
    AboveThreshold is $(\eps, 0)$-DP.
\end{lemma}

Within the proof of Theorem 3.25 of their work (which is not exactly what we want), the authors use the Laplace tail bound and a union bound to show the following statement.
\begin{lemma}
\label{lem:above-utility}
Let
\[
\alpha = \frac{8(\log k + \log(2/\beta))}{\eps}.
\]
Then, with probability $\beta$, the following holds across all timesteps:
\begin{itemize}
    \item If $a_i = \yes$, then $f_i(D) \geq T - \alpha$.
    \item If $a_i = \no$, then $f_i(D) \leq T + \alpha$.
\end{itemize}
\end{lemma}

Finally, for our proof, we need the advanced composition theorem, appearing as Corollary 3.21 in their work.
\begin{theorem}\label{thm:adv-composition}
   For all $\eps \in (0,1)$ and $\delta > 0$, $k$-fold adaptive composition of $(\eps', 0)$-DP mechanisms satisfies $(\eps, \delta)$-DP as long as 
   \[
    \eps' \leq \frac{\eps}{2\sqrt{2k\ln(1/\delta)}}
   \]
\end{theorem}
Now, we are ready to prove the version of SVT theorem we use.
\svt*

\begin{proof}
We implement SVT by composition of at most $c$ calls to AboveThreshold.
We normalize and recenter all queries $q_i$ for SVT to $f_i = (q_i - T_i) / \Delta$ for AboveThreshold with a fixed threshold $T=0$.
When a call to AboveThreshold returns $\no$, SVT will return $\no$.
When a call to AboveThreshold returns $\yes$ and halts, SVT will return $\yes$ and instantiate a fresh call to AboveThreshold if the number of $\yes$ outputs is less than $c$.
Otherwise, SVT will return $\halt$ for the remainder of time.

Let  $\eps' = \frac{\eps}{2\sqrt{2c\ln(1/\delta)}}.$ Note that SVT's only interaction with the private dataset $D$ is via the outputs of at most $c$ instantiations of AboveThreshold.
Using $\eps'$ as the privacy parameter for each call to AboveThreshold, SVT is $(\eps, \delta)$-DP by \cref{lem:above-dp,thm:adv-composition}.

The utility of SVT follows from that of AboveThreshold via \cref{lem:above-utility} plugging in $\eps'$ and $\beta' = \beta/c$ in order to union bound over the failure probability of each call to AboveThreshold.
\cref{lem:above-utility} gives additive error
\[
    \alpha = \frac{8(\log k + \log(2/\beta'))}{\eps'} 
    = \frac{16(\log k + \log(2c/\beta)) \sqrt{2c\ln(1/\delta)}}{\eps}.
\]
Since AboveThreshold operates on the normalized queries $f_i = q_i/\Delta$, the additive error term in the final theorem statement has an additional factor $\Delta$.
\end{proof}

\end{document}